\documentclass[]{JHEP3}

\title{Poincare covariant mechanics on noncommutative space.}
\author{A. A. Deriglazov\\ Departamento de Matematica,
Universidade Federal de Juiz de Fora, Juiz de Fora, MG, Brasil\\ 
E-mail: \email{alexei@ice.ufjf.br}
~\footnote{on leave of absence from 
Dept. Math. Phys., Tomsk Polytechnical University, Tomsk, Russia}
}
\abstract{
The Dirac approach to constrained systems can be adapted to construct relativistic 
invariant theories on a noncommutative (NC) space. As an example,  
we propose and discuss relativistic invariant NC particle 
coupled to electromagnetic (EM) field by means of the standard term 
$A_{\mu}\dot x^{\mu}$. 
Poincare invariance implies deformation of the free particle NC algebra in the 
interaction theory. The corresponding corrections survive in the nonrelativistic limit.}
\preprint{}
\begin{document}
\section{Introduction: noncommutative geometry and relativistic invariance.}

Noncommutativity of space coordinates [1, 2] can be naturally incorporated into the 
quantum field theory framework, at least for mechanical models [3-8], using the  
Dirac approach to constrained systems [9-11] appropriately adopted to the case. 
The method implies that one works with the 
Lagrangian action in the first order form, the latter being characterized 
by doubling of the configuration space variables: $q^A\longrightarrow (q^A, v_A)$. Noncommutative (NC) 
version of an arbitrary nondegenerate 
mechanics can be obtained [7] by addition of a Chern-Simons type term for $v_A$ to the 
first order action of the initial system:  
$S_{NC}=S_1(q, v)+\int d\tau\dot v_A\theta^{AB}v_B$. 
Nontrivial bracket for the 
position variables $q^A$ appears in this case as the Dirac bracket,
after taking into account the constraints present in the model.
The numerical matrix 
$\theta^{AB}=-\theta^{BA}$, originating 
from the Chern-Simons term, then becames a NC parameter of the formulation.  
Quantization of the NC system leads to quantum
mechanics with ordinary product replaced by the Moyal product.
Various new aspects arising in classical and quantum mechanics on NC space were 
observed (see [12-20] and references therein). 
An apparent defect of the NC models is a lack of relativistic invariance, mainly 
due to the fact that the noncommutativity parameter is a constant matrix.

In this work we demonstrate that the noncommutativity does not violate 
necessarily the Lorentz invariance, but does lead to non standard realization of 
the Lorentz group, at least for the particle models. 
In the abovementioned approach, coordinates of the extended configuration 
space are ordinary (commuting) quantities. 
So, there is no principal difficulty to construct a relativistic invariant theory on 
the extended space. After that, 
taking into account the constraints present in the model, one obtains a theory in 
physical subspace, the latter being noncommutative and endowed automatically with the 
Poincare structure. In particular, {\em{NC brackets of the variables are 
covariant under the Poincare transformations}}. 

Let us discuss how this scheme can be realized for a   
$D$-dimensional particle (see [8] for details). 
Chern-Simons term can
be added to the first order action of the relativistic particle, without 
spoiling the reparametrization invariance. As a consequence,
the model will contain the desired relativistic constraint $p^2-m^2=0$.
The problem is
that the numerical matrix $\theta^{\mu\nu}$ does not respect the Lorentz
invariance. To resolve the problem, let us consider a particle interacting
with a new configuration-space variable
$\theta^{\mu\nu}(\tau)=-\theta^{\nu\mu}(\tau)$,
instead of the constant matrix. An appropriate action is 
\begin{eqnarray}\label{1}
S_0=\int d\tau\left[\dot x^\mu v_\mu-\frac{e}{2}(v^2-m^2)+
\frac{1}{\theta^{\alpha\beta}\theta_{\alpha\beta}}
\dot v_\mu\theta^{\mu\nu}v_\nu\right],
\end{eqnarray}
Here $x^\mu(\tau), ~ v^\mu(\tau), ~ e(\tau), ~ \theta^{\mu\nu}(\tau)$,
are the configuration space variables of the model,
$\eta^{\mu\nu}=(+,-, \ldots ,-)$. Insertion of the term $\theta^2$
in the denominator has the same meaning as for the eibein in
the action of massless particle: $L=\frac{1}{2e}\dot x^2$. Technically,
it rules out the degenerate gauge $e=0$. 
The action is manifestly
invariant under the Poincare transformations 
$\delta x^\mu=\omega^\mu{}_\nu x^\nu+a^\mu$, 
$\delta v^\mu=\omega^\mu{}_\nu v^\nu$, 
$\delta\theta^{\mu\nu}=\omega^{[\mu}{}_\alpha\theta^{\alpha\nu]}$. 
The crucial observation is that the action is invariant also under the following 
local symmetry 
(with the parameter $\epsilon_{\mu\nu}(\tau)=-\epsilon_{\nu\mu}(\tau)$)
\begin{eqnarray}\label{2}
\delta x^\mu=-\epsilon^{\mu\nu}v_\nu, \qquad
\delta\theta_{\mu\nu}=-(\theta^{\alpha\beta}\theta_{\alpha\beta})\epsilon_{\mu\nu}+
2\theta_{\mu\nu}(\theta^{\alpha\beta}\epsilon_{\alpha\beta}),
\end{eqnarray}
associated with the variable $\theta$. 
The latter can be gauged out (the corresponding first class constraints 
will be discussed below), an admissible gauge is
\begin{eqnarray}\label{3}
\theta^{0i}=0, \qquad \theta^{ij}=const, \qquad 
\theta^2\equiv\theta_{ij}\theta_{ji}\ne 0. 
\end{eqnarray}
The situation here is similar to the string theory, where the  
two-dimensional metric $g_{ab}(\sigma^c)$ can be gauged out completely (using the 
reparametrization symmetry): $g_{ab}=\eta_{ab}$. 
The noncommutativity parameter of the
gauge fixed version is the numerical matrix $\theta^{ij}$.
The model is also subject to second class constraints 
$G^\mu\equiv p^\mu-v^\mu=0, ~ 
T^\mu\equiv\pi^\mu-\frac{1}{\theta^2}\theta^{\mu\nu}v_\nu=0$, 
where $p, \pi$ are conjugated momenta for $x, v$.
Let us construct the corresponding Dirac bracket. 
After that, taking $x, p$ as the physical sector variables, remaining variables can 
be omitted from consideration by using the constraints. The physical subspace 
turns out to be noncommutative, with the following Dirac brackets for the physical 
variables: 
\begin{eqnarray}\label{4}
\{x^\mu, x^\nu\}=2\theta^{-2}\theta^{\mu\nu}, \quad
\{x^\mu, p_\nu\}=\delta^\mu_\nu,
\quad \{p_\mu, p_\nu\}=0,
\end{eqnarray}
where Eq.(\ref{3}) is implied. Dynamics of the 
resulting theory corresponds to the free particle, the Hamiltonian equations of 
motion are 
$\dot x^\mu=ep^\mu, ~ \dot p^\mu=0$. They are accompanied by the
remaining first class constraint $p^2-m^2=0$. 
As it usually happens in a
theory with local symmetries, Poincare invariance of the gauge
fixed version is combination of the initial Poincare and local
transformations which preserve the gauge (\ref{3}) 
\begin{eqnarray}\label{5}
\delta x^\mu=\omega^\mu{}_\nu x^\nu-
\epsilon^{\mu\nu}(\omega) v_\nu+a^\mu, \quad
\delta v^\mu=\omega^\mu{}_\nu v^\nu; \cr 
\epsilon^{\mu\nu}(\omega)\equiv -
\theta^{-2}\omega^{[\mu}{}_\rho\theta^{\rho\nu]},
\end{eqnarray}
where Eq.(\ref{3}) for $\theta^{\mu\nu}$ is implied. By construction, 
{\em{the brackets (\ref{4}) are covariant with 
respect to the Poincare transformations (\ref{5})}} (note that $v=p$ on the constraint 
surface). In the interacting theory Poincare invariance will require deformation 
of the free NC algebra, see (\ref{11}), (\ref{15}). 

Of course, the free particle is not very interesting object 
(similarly to the nonrelativistic case, any traces of noncommutativity disappear 
when formulated in terms of canonical variables
$\hat x^\mu\equiv x^\mu+\frac{1}{\theta^2}\theta^{\mu\nu}v_\nu$,
see also (\ref{1})). A possibility to introduce some rather exotic interaction with 
electromagnetic (EM) field was discussed previously in [21]).
Below we demonstrate that NC relativistic particle couples 
to an external EM field by means of the standard term: $A_\mu\dot x^\mu$. 

\section{Partially gauge fixed action of NC relativistic particle and 
interaction with the electromagnetic field.}

To construct action for NC particle on EM background in the 
manifestly Poincare invariant 
form one needs to ensure, besides the Poincare symmetry, 
the local $\epsilon$-symmetry 
(\ref{2}) and $U(1)$ gauge symmetry (another technically nontrivial task 
is to preserve the first class character of the 
relativistic  constraint $p^2-m^2=0$). 
To simplify the problem, we consider here partially gauge fixed version of 
the model which arises after fixing the gauge (\ref{3}) for $\epsilon$-symmetry . 
In this case, one needs to ensure the Poincare 
and $U(1)$ symmetry of the action, which is a more simple task. 

Starting from manifestly Poincare invariant action of NC particle (1) 
one finds in the Hamiltonian formulation the following constraints 
\begin{eqnarray}\label{99}
G^\mu\equiv p^\mu-v^\mu=0, \qquad \qquad
T^\mu\equiv\pi^\mu-\frac{1}{\theta^2}\theta^{\mu\nu}v_\nu=0, \cr
v^2-m^2=0, \qquad
p_{\theta}^{\mu\nu}=0, \qquad  p_e=0.
\end{eqnarray}
Poisson brackets of the constraints are 
\begin{eqnarray}\label{100}
\{G^\mu, G^\nu\}=0, \qquad \{T^\mu, T^\nu\}=
-\frac{2}{\theta^2}\theta^{\mu\nu}, \qquad
\{G_\mu, T^\nu\}=-\delta_\mu^\nu, \cr
\{v^2-m^2, T^\mu\}=2v^\mu, \qquad
\{T_\mu, p_\theta^{\rho\sigma}\}=-\frac{1}{\theta^2}
\delta_\mu^{[\rho}v^{\sigma ]}+
\frac{4}{\theta^4}(\theta v)_\mu\theta^{\rho\sigma}.
\end{eqnarray}
Using Eq.(\ref{100}) one can separate the first class constraints related with 
$\theta^{\mu\nu}$ variable 
\begin{eqnarray}\label{101}
P_\theta^{\rho\sigma}=p_\theta^{\rho\sigma}-
G^\alpha\{T_\alpha, p_\theta^{\rho\sigma}\}=0.
\end{eqnarray}
$P_\theta^{\rho\sigma}$ has vanishing brackets with other 
constraints and corresponds 
to the local symmetry (2). 
In the gauge (3) for the constraints (\ref{101}) only the space components 
of the variable $\theta^{\mu\nu}$ survive. 
The corresponding partially gauge fixed version of (1) is, evidently 
\begin{eqnarray}\label{6}
S_{gf}=\int d\tau\left[\dot x^\mu v_\mu-\frac{e}{2}(v^\mu v_\nu-m^2)-
\frac{1}{\theta^2}\dot v_i\theta^{ij}v_j\right],
\end{eqnarray}
where $\theta_{ij}$ {\em{is now the numerical matrix from (\ref{3})}}. Physical 
sector dynamics of the 
models (\ref{1}) and (\ref{6}) is the same, which can be confirmed by 
direct computations. 

Let us discuss interaction with the EM background. 
In the gauge fixed formulation (\ref{6}), the Poincare transformation (\ref{5}) for 
the variable $x^\mu$ involves $v^\mu$. For this reason the 
background field $A^\mu(x)$ 
is not a Poincare covariant quantity. One can take a field with 
general dependence on the configuration space variables: $A^\mu(x, v)$. 
With this choice, equations of motion will have corrections of order 
$O(\frac{\theta}{\theta^2})$ as well as other deviations from commutative case  
involving  
$\frac{\partial A_\mu}{\partial v^\nu}$. Since it is not a theory 
which one expects as 
NC version of ordinary particle, we consider 
here another possibility. 
Namely, from Eq.(\ref{5}) one finds Poincare invariant subspace parametrised 
by the coordinates
\begin{eqnarray}\label{7}
\hat x^\mu\equiv x^\mu+\frac{1}{\theta^2}\theta^{\mu\nu}v_\nu,
\end{eqnarray}
which obey $\delta_\omega\hat x^\mu=\omega^\mu{}_\nu\hat x^\nu$. 
(It is interesting to note that they  
coincide with the canonical coordinates of the free theory).
The field $A^\mu(\hat x)$ is a Poincare covariant object. So, as the interaction 
theory, let us consider the following action 
\begin{eqnarray}\label{8}
S=\int d\tau\left[\dot x^\mu v_\mu-\frac{e}{2}(v^\mu v_\mu-m^2)-
\frac{1}{\theta^2}\dot v_i\theta^{ij}v_j+\right.\cr
\left.\dot x^\mu A_\mu\left(x+\frac{\theta v}{\theta^2}\right)\right].
\end{eqnarray}
Note that $\theta$-dependence can not be removed now by a change of variables, 
similarly to the case of NC quantum mechanics with a potential.  
Symmetries of the action are reparametrizations, Poincare symmetry  
and local gauge symmetry associated with $A^\mu(\hat x)$. 
The Poincare transformations are represented by Eq.(\ref{5}), accompanied by   
$\delta A^\mu=\omega^\mu{}_\nu A^\nu-$
$n_\mu(\dot v\epsilon(\omega)A)$. 
Here $n^\mu\equiv\frac{\dot x^\mu}{(\dot x)^2}$ is a unit vector 
in the direction of 
motion of the particle \footnote{Appearance of velocity dependent 
terms in the Poincare transformation law 
is a common property of gauge fixed versions. For example, the transformations for 
ordinary particle in the light-cone gauge are of the form 
$\delta x^\mu=\omega^\mu{}_\nu x^\nu-(\omega^0{}_\nu x^\nu)\dot x^\mu$, ~  
$\delta v^\mu=\omega^\mu{}_\nu v^\nu-(\omega^0{}_\nu x^\nu)\dot v^\mu$, ~  
$\delta A^\mu=\omega^\mu{}_\nu A^\nu-(\omega^0{}_\alpha x^\alpha)\dot x^\nu
\partial_\nu A_\mu$.}.
The transformation is non singular on-shell. 

Since the background field depends both on $x^\mu$ and 
$v^\mu$, one expects that $U(1)$ symmetry of the gauge fixed version will have a 
corrected form (note that the symmetry $\delta A_\mu=\partial_\mu\alpha(x)$ 
is not sufficient to 
gauge away the time-like component $A_0(\hat x)$). It is indeed the case, and  
local gauge symmetry of (\ref{8}) has the form 
\begin{eqnarray}\label{9}
\delta A_\mu(\hat x)=\partial_\mu\alpha(\hat x)-
\theta^{-2}n_\mu(\dot v\theta\partial\alpha(\hat x)), 
\end{eqnarray}
with the parameter $\alpha(\hat x)$. Here and below, the derivative $\partial_\mu$ 
is calculated with respect 
to $\hat x^\mu$. As in the commutative space, the symmetry (\ref{9}) 
ensures absence of negative norm modes of EM field. 

\section{Hamiltonian analysis of the model.}

To analyze physical sector of the constrained system (\ref{8}), we rewrite
it in the Hamiltonian form. The interaction term leads to a nonlinear 
deformation of the constraint structure as compared with the free case. In all 
the equations below  
$\theta^{\mu\nu}$ is the numerical matrix (\ref{3}). The following notations 
will be used: 
\begin{eqnarray}\label{10}
\hat\theta^{\mu\nu}\equiv 2\theta^{-2}\theta^{\mu\nu}, \quad 
\triangle_{\mu\nu}\equiv\eta_{\mu\nu}-
\theta^{-2}\theta_\mu{}^\alpha\partial_\alpha A_\nu, \cr 
D\equiv\left[\triangle+\hat\theta(\triangle^T)^{-1}F\right]^{-1},
\end{eqnarray}
where $F_{\mu\nu}$ is the field strength.
Starting from the action (\ref{8}), one finds in the Hamiltonian
formalism the primary constraints
\begin{eqnarray}\label{102}
G^\mu\equiv p^\mu-v^\mu-A_\mu(\hat x)=0, \qquad 
T^\mu\equiv\pi^\mu+\frac{1}{\theta^2}\theta^{\mu\nu}v_\nu=0, \cr  
p_e=0.
\end{eqnarray}
and the Hamiltonian
\begin{eqnarray}\label{123}
H=\frac{e}{2}(v^\mu v_\mu-m^2)+\lambda_{1\mu}G^\mu+\lambda_{2\mu}T^\mu+
\lambda_ep_e.
\end{eqnarray}
Here $p, ~ \pi$ are conjugated momentum for $x, ~ v$ and $\lambda$ are the
Lagrangian multipliers for the constraints. On the next step there is
appear the secondary constraint
\begin{eqnarray}\label{124}
v^\mu v_\mu-m^2=0,
\end{eqnarray}
as well as equations for determining the Lagrangian multipliers
\begin{eqnarray}\label{125}
\lambda_1=eDv, \qquad
\lambda_2=e(\triangle^T)^{-1}FDv.
\end{eqnarray}
There is no of tertiary constraints in the problem.
Poisson brackets of the constraints are
\begin{eqnarray}\label{126}
\{G^\mu, G^\nu\}=F^{\mu\nu}, \quad \{T^\mu, T^\nu\}=
\frac{2}{\theta^2}\theta^{\mu\nu}, \quad
\{T_\mu, G_\nu\}=\triangle_{\mu\nu}. \quad
\end{eqnarray}
The constraints $G^\mu, ~ T^\mu$ form the second class system and can
be taken into account by transition to the Dirac bracket. It´s manifest form is
\begin{eqnarray}\label{127}
\{A, B\}_D=\{A, B\}-\{A, G_\mu\}
\left[D\hat\theta\left(\triangle^T\right)^{-1}\right]^{\mu\nu}
\{G_\nu, B\}- \cr
\{A, G_\mu\}D^{\mu\nu}\{T_\nu, B\}+ 
\{B, G_\mu\}D^{\mu\nu}\{T_\nu, A\}- \cr
\{A, T_\mu\}\left[\left(\triangle^T\right)^{-1}FD\right]^{\mu\nu}\{T_\nu, B\}.
\end{eqnarray}
Now the constraints (\ref{102}) can be used to omit part of variables 
from consideration. 
Since the constraint $G^\mu=0$ can not be resolved with respect to $v^\mu$, we take 
the variables $(x, v)$ as the physical sector ones\footnote{First order formulation 
implies that one can choose different pairs of variables as the physical ones [5].   
It leads to equivalent descriptions of the system, as it should be. 
For example, for ordinary particle on EM background the sectors $x, p$ and $x, v$ 
are related as follows: $p^\mu=v^\mu+A^\mu$.}. 
The Dirac brackets (in the matrix notations) for the physical variables are  
\begin{eqnarray}\label{11}
\{x, x\}=D\hat\theta(\triangle^T)^{-1}, \quad
\{x, v\}=D, \cr
\{v, v\}=(\triangle^T)^{-1}FD.
\end{eqnarray}
In the decoupling limit $A^\mu\to 0$ one has the free particle NC algebra (\ref{4}). 
The Hamiltonian equations of motion for the theory are 
\begin{eqnarray}\label{12}
\dot x=eDv, \qquad  
\dot v=e(\triangle^T)^{-1}FDv.
\end{eqnarray}
They are accompanied by the constraint $v^2-m^2=0$.  
The equations (\ref{12}) imply the following second order equation for the 
coordinate of the particle (in the gauge $e=1$):   
\begin{eqnarray}\label{13}
\ddot x=D\left[(\triangle^T)^{-1}F-
(D^{-1})^{.}\right]\dot x. 
\end{eqnarray}
Poincare invariant dynamics on NC space acquires nontrivial 
corrections of order $\hat\theta$, as compared with the ordinary particle. 
Expanding r.h.s. of Eqs. (\ref{11})-(\ref{13}) up to $O^2(\hat\theta)$, 
one obtains 
the first order corrections as follows:
\begin{eqnarray}\label{14}
\dot x^\mu=ev^\mu-e\left[\hat\theta(F-\frac{1}{2}\partial A)\right]
^{\mu\nu}v_\nu, \cr
\dot v^\mu=eF_{\mu\nu}v^\nu+e\left[\frac{1}{2}(\hat\theta v)
^\alpha\partial_\alpha F_{\mu\nu}-
(F\hat\theta F)_{\mu\nu}\right.+ \cr 
\left.\frac{1}{2}(F\hat\theta\partial A) _{[\mu\nu]}\right]v^\nu.
\end{eqnarray}
\begin{eqnarray}\label{15}
\{x^\mu, x^\nu\}=\hat\theta^{\mu\nu}, \cr 
\{x^\mu, v_\nu\}=\delta^\mu_\nu-\hat\theta^{\mu\alpha}
(F-\frac{1}{2}\partial A)_{\alpha\nu}, \cr 
\{v_\mu, v_\nu\}=F_{\mu\nu}+
\frac{1}{2}(\hat\theta v)^\alpha\partial_\alpha F_{\mu\nu}- \cr 
(F\hat\theta F)_{\mu\nu}+
\frac{1}{2}(F\hat\theta\partial A) _{[\mu\nu]}.
\end{eqnarray}
\begin{eqnarray}\label{16}
\ddot x=F\dot x+\frac{1}{2}(\hat\theta v)^\alpha\partial_\alpha F\dot x
-\hat\theta FF\dot x+ \cr
\frac{1}{2}\hat\theta\partial AF\dot x- 
\frac{1}{2}\dot xF\hat\theta\partial A-
\hat\theta(F-\frac{1}{2}\partial A)^{.}\dot x.
\end{eqnarray}
In the commutative limit $\hat\theta\to 0$ the equations 
(\ref{14})-(\ref{16}) turn out into the corresponding equations for 
ordinary particle 
on the background $A^\mu(x)$. In this limit the Poincare transformations 
acquire the standard form (the variable $v^\mu$ disappears 
from the transformation
law for $x^\mu$). 

\section{Conclusion.}
 
In this work we have presented D-dimensional NC relativistic particle coupled  
to the background EM field. To construct Poincare invariant theory on 
NC space one needs 
to replace the numerical NC parameter $\theta^{\mu\nu}$ by a dynamical variable. 
Local $\epsilon$-symmetry associated with the variable allows one to impose 
the gauge $\theta^{0i}=0, \theta^{ij}=const$. The numerical matrix
$\theta^{ij}$ turns out to be a NC parameter of the physical subspace. 
The standard interaction term can be added to the 
partially gauge fixed action for the free NC particle (\ref{6}) without spoiling 
the Poincare symmetry or the $U(1)$ gauge symmetry. Poincare invariance on the 
NC space requires special dependence of the background field on the 
variables of the first order formulation: $A^\mu(x+\frac{\theta v}{\theta^2})$  
(for our choice of the physical variables, 
they turn out into phase space variables in the Hamiltonian formulation). 
In turn, this implies modified form of the $U(1)$-transformations (\ref{9}). 
Another consequence of this special dependence is that the 
Poincare invariant dynamics on NC space acquires higher nonlinear corrections 
of order $\hat\theta$ as compared with ordinary particle. Let us point that 
the method developed can be directly applied to investigate the rotational 
invariance in the NC quantum mechanics as well.   
 
The equations (\ref{14})-(\ref{16}) can be a starting point for 
various computations of 
leading corrections to ordinary (commutative) theory. An important conclusion 
from Eq.(\ref{15})  
is that in an interacting Poincare invariant theory one can not use the free 
particle NC algebra (\ref{4}) as the leading approximation in $\hat\theta$, since 
the latter is given by (\ref{15}). Nonrelativistic limit of the algebra (\ref{15}) 
gives the corresponding NC quantum mechanics with brackets which are different 
from (\ref{4}). It can imply revision of the results obtained 
in NC quantum mechanics 
on the basis of Eq.(\ref{4}) [15, 19, 20]. 
To analyze these problems, it will be interesting to 
find a set of canonical variables (i.e. the ones with canonical brackets instead 
of Eq.(\ref{15}). Another interesting problem is to find a manifestly 
Poincare invariant version of the model (\ref{8}). We suppose also that that the 
scheme can be realised for the open string on a $B$-field 
background. The formulation 
developed in [6] in terms of infinitely dimensional mechanics seems to be 
appropriate for this aim.
These problems will be discussed elsewhere.

\end{document}